\pdfoutput=1
\documentclass[utf8]{FrontiersinHarvard}
\usepackage{url,hyperref,lineno,microtype,subcaption}
\usepackage[onehalfspacing]{setspace}
\usepackage{natbib}
\usepackage{doi}  
\usepackage{hyperref}  
\hypersetup{
    colorlinks=true,
    linkcolor=black,
    citecolor=darkgray,
    urlcolor=black
}
\usepackage{comment} 
\usepackage{enumerate} 
\usepackage{graphicx}
\usepackage{csvsimple}
\usepackage{makecell}
\usepackage[most]{tcolorbox}
\usepackage{colortbl}
\definecolor{gray}{rgb}{0.9,0.9,0.9} 

\usepackage{algorithm}
\usepackage{algpseudocode}

\newtcolorbox{modifiedbox}[1][]{%
  colback=yellow, 
  colframe=yellow, 
  boxrule=0pt, 
  boxsep=0pt, 
  left=2pt, 
  right=2pt, 
  top=2pt, 
  bottom=2pt, 
  sharp corners,
  before skip=0pt,
  after skip=0pt, 
  #1
}
\newcommand{\argmin}{\mathop{\mathrm{arg\,min}}\limits}

\def\keyFont{\fontsize{8}{11}\helveticabold }
\def\firstAuthorLast{Mikiya Doi {et~al.}} 
\def\Authors{Mikiya Doi\,$^{1,*}$, Yoshihiro Nakao\,$^{2}$ , Takuro Tanaka \,$^{2}$, Masami Sako\,$^{2}$ and Masayuki Ohzeki\,$^{1,3,4,5}$} 
\def\Address{$^{1}$ Graduate School of Information Sciences, Tohoku University, Sendai 980-8579, Japan\\
$^{2}$ LG Japan Lab Ltd., Yokohama, Kanagawawa 220-0011, Japan \\
$^{3}$ International Research Frontier Initiative, Tokyo Institute of Technology, Tokyo, 108-0023, Japan \\
$^{4}$ Department of Physics, Tokyo Institute of Technology, Meguro, Tokyo 152-8550, Japan \\
$^{5}$ Sigma-i Co. Ltd., Minato, Tokyo 108-0075, Japan
}

\def\corrAuthor{Mikiya Doi}
\def\corrEmail{mikiya.doi.t2@dc.tohoku.ac.jp}

\begin{document}

\onecolumn

\firstpage{1}

\title[Black-box optimization using quantum annealer]{Exploration of new chemical materials using black-box optimization with the D-wave quantum annealer} 

\author[\firstAuthorLast ]{\Authors} 
\address{} 
\correspondance{} 

\extraAuth{}

\maketitle

\begin{abstract}

In materials informatics, searching for chemical materials with desired properties is challenging due to the vastness of the chemical space. 
Moreover, the high cost of evaluating properties necessitates a search with a few clues. In practice, there is also a demand for proposing compositions that are easily synthesizable.
In the real world, such as in the exploration of chemical materials, it is common to encounter problems targeting black-box objective functions where formalizing the objective function in explicit form is challenging, and the evaluation cost is high. In recent research, a Bayesian optimization method has been proposed to formulate the quadratic unconstrained binary optimization (QUBO) problem as a surrogate model for black-box objective functions with discrete variables. Regarding this method, studies have been conducted using the D-Wave quantum annealer to optimize the acquisition function, which is based on the surrogate model and determines the next exploration point for the black-box objective function.
In this paper, we address optimizing a black-box objective function containing discrete variables in the context of actual chemical material exploration. 
In this optimization problem, we demonstrate results obtaining parameters of the acquisition function by sampling from a probability distribution with variance can explore the solution space more extensively than in the case of no variance. As a result, we found combinations of substituents in compositions with the desired properties, which could only be discovered when we set an appropriate variance.

\tiny
 \keyFont{ \section{Keywords:} quantum annealing, quantum computing, black-box-optimization, combinatorial optimization problem, materials informatics} 
\end{abstract}

\section{Introduction}
\label{sec: intro}

Black-box optimization is a method to optimize a function that does not have an explicit objective function in the mathematical form. In the real world, this optimization problem appears in various fields, including material informatics, robotics \citep{deisenroth2011SurveyFNTinRobotics}, machine learning \citep{snoek2012PracticalAdv.NeuralInf.Process.Syst.}, and recommendation systems \citep{Vanchinathan2014}. 
Bayesian optimization is one of the solutions for black-box optimization problems \citep{jones1998EfficientJournalofGlobalOptimization}.
Taking the exploration of chemical materials as an example, a surrogate model is constructed using an existing dataset to predict the relationship between the combinations of substituents in the chemical materials and the corresponding property values. Based on this surrogate model, an acquisition function is defined. The combination of substituents obtained through optimizing this acquisition function is then used as the next input point for the black-box objective function, enabling the evaluation of the actual property values. 
The relationship between the inputted combination of substituents and the actual property value is then added to the existing dataset, then updating the surrogate model. Repeating this process is to explore the combinations of substituents that yield the desired property values.
Especially for black-box optimization problems involving discrete variables, discrete variables are included in both the surrogate model and the acquisition function. Therefore, even optimizing the acquisition function often proves to be NP-hard, and the solutions obtained through optimization are generally approximate. In a previous study, Bayesian optimization of combinatorial structures (BOCS) was proposed as the promising algorithm for such problems \citep{baptista2018BayesianProc.35thInt.Conf.Mach.Learn.a}. In this algorithm, the acquisition function was assumed as quadratic unconstrained binary optimization (QUBO) problem.

Quantum annealing \citep{kadowaki1998QuantumPhys.Rev.E} is a heuristic algorithm to solve QUBO problems by driving binary variables through quantum fluctuations. Many well-known combinatorial optimization problems can be encoded into QUBO problems \citep{Lucas2014}. Practical applications of quantum annealing can be found in various fields, including traffic flow optimization \citep{neukart2017TrafficFront.ICT, Inoue2021, shikanai2023Traffic}, manufacturing \citep{Ohzeki2019, habaTravelTimeOptimization2022}, finance \citep{Rosenberg2015, venturelli2019ReverseQuantumMach.Intell.}, steel manufacturing \citep{Yonaga2022}, decoding problems \citep{ide2020Maximum2020Int.Symp.Inf.TheoryItsAppl.ISITA, arai2021MeanPhys.Rev.Res.}, and algorithms in machine learning \citep{Amin2018, OMalley2018, urushibata2022ComparingJ.Phys.Soc.Jpn., hasegawa2023Kernel, goto2023Online}.
Furthermore, quantum annealing, which utilizes the quantum tunneling effect, is expected to find the optimal solution for several combinatorial optimization problems more rapidly than algorithms such as simulated annealing \citep{Kirkpatrick1983}. This advantage is investigated from the perspective of energy landscape characteristics \citep{das2008ColloquiumRev.Mod.Phys.} and through numerical computation \citep{denchev2016WhatPhys.Rev.X}. In addition, there are discussions about the characteristics of solutions obtained in cases where multiple optimal solutions exist \citep{Yamamoto2020, maruyama2021Graph}. With these backgrounds, quantum annealing has recently attracted attention, both for its potential applications and for validating the fundamental aspects of quantum effects.

Studies that employ quantum annealing in some algorithms for black-box optimization problems involving discrete variables exist. 
These include benchmark tests \citep{koshikawa2021BenchmarkJ.Phys.Soc.Jpn.} that have examined the presence or absence of quantum superiority in optimizing acquisition functions. 
In terms of practical applications, there are case studies that have achieved significant screening in the exploration of chemical materials within the search chemical space \citep{hatakeyama-sato2021TacklingAdvancedIntelligentSystems, takurotanaka2023VirtualJ.Phys.Soc.Jpn.}, as well as 
instances of designing complex metamaterials \citep{kitai2020DesigningPhys.Rev.Research}.

In the exploration of chemical materials, it is necessary not only to discover molecules with the desired property values but also to be concerned about scenarios in actual synthesis where molecules with specific substructures may become entirely unfeasible to synthesize. Drawing inspiration from previous studies and practical needs, we demonstrate a method for proposing diverse compositions of chemical materials with desired properties, targeting a black-box optimization problem that includes discrete variables in actual chemical material exploration. In more detail, we show results that by obtaining parameters of the surrogate model and acquisition function from sampling a probability distribution with an appropriate variance and optimizing the acquisition function, we explored the solution space more extensively while optimizing the black-box objective function. The method used in this paper is generally referred to as Thompson sampling \citep{THOMPSON1933, NIPS2011_e53a0a29}. In this sense, it can be said that our research results evaluate the impact of the magnitude of the variance of the posterior probability distribution in Thompson sampling.

The remaining sections of this paper are organized as follows:
In the next section, Section 2, we explain the problem setting in this paper and the method we propose. In Section 3, we demonstrate the results of the experiments related to the actual exploration of chemical materials. Finally, Section 4 summarizes our research and discusses this paper and future research directions.
\section{Materials and Methods}
\label{sec: materials and method}
In this section, we introduce the problem settings based on the search for chemical materials, which is the focus of this paper. 
Subsequently, in Bayesian optimization, we explain the construction of the surrogate model in the QUBO form, which is well-known in prior research, along with the construction of the acquisition function. We provide this explanation in conjunction with our method aim.

\subsection{Problem settings}
\label{subsec: problem settings}
In this paper, we define the binding of substituents to specific sites of the molecular frame as the composition of chemical materials. 
We aim to propose various combinations of substituents through Bayesian optimization while maximizing a target material property value. To align our description with other literature focusing on black-box optimization problems, we define our goal as a minimization problem, utilizing the fact that maximization and minimization problems can be transformed into each other by reversing the sign of the objective function. 

\subsection{Methods} 
\label{subsubsec: method}
We express the assignment of substituents using a binary vector. In particular, for substituents that can bind to each site, we encode them by converting the 0-indexed substituent number to binary. 
Thus, we set a binary vector $\vec{x}^{(\mu)} \in \{0, 1\}^{N}$ as input, and the corresponding target material property value $y^{(\mu)}$ as output.
We aim to find $\vec{x}$ that minimizes a black-box objective function. Since we cannot know an explicit form of the black-box objective function, we construct a surrogate model as QUBO form following the previous studies. We utilize an existing dataset $\mathcal{D} = \{ \vec{x}^{(\mu)}, y^{(\mu)} \}_{\mu = 1}^{D}$ to sample the parameters of the surrogate model from a probability distribution we discuss later and construct it. Based on the surrogate model, we construct an acquisition function and propose a combination of substituents that optimize the acquisition function using the D-Wave quantum annealer. Subsequently, we input the proposed combination of substituents as the next exploration point $\vec{x}^{(new)}$ and obtain output $y^{(new)}$ from the black-box objective function. Then we append $\{ \vec{x}^{(new)}, y^{(new)} \}$ to the existing dataset as new data and reconstruct the surrogate model. By repeating this process, we aim to obtain diverse combinations of substituents with desired target material property values.
\subsubsection{Construction of surrogate model function}

\label{subsubsec: surrogate model}
We construct the surrogate model $f_{surrogate}(\vec{x})$ in the QUBO form in this paper. 
\begin{equation}
f_{surrogate} (\vec{x}) = \alpha_{0} + \sum_{i} \alpha_{i} x_{i} + \sum_{i < j}  \alpha_{ij} x_{i} x_{j} 
\label{eq:01}
\end{equation}
For simplicity, we set the surrogate model parameters $\{\alpha_{i}, \alpha_{ij}\} = \vec{\alpha} \in \mathbb{R}^{p}$. Note that $p=1+N+N(N-1)/2$.
Defining $X \in \{0, 1\}^{D \times p}$ as the design matrix and denoting the $\mu$-th row in the design matrix $X$ as $X^{(\mu)}$, we have the following expression
$X^{(\mu)} = \left(1, x^{(\mu)}_{1}, ..., x^{(\mu)}_{N}, \, x^{(\mu)}_{1}x^{(\mu)}_{2}, x^{(\mu)}_{1}x^{(\mu)}_{3}, ...,
x^{(\mu)}_{N-1}x^{(\mu)}_{N}\right)$. Furthermore, we set the output vector $\vec{y} \in \mathbb{R}^{D}$ and $I$ as the identity matrix. Then, we assume a prior distribution of surrogate model parameters $P(\vec{\alpha})$ with a variance $\sigma^{2}_{\alpha}I$ and a likelihood function over the surrogate model parameters $\vec{\alpha}$ with a variance $\sigma^{2}_{y}I$. We give the prior distribution and likelihood function as following multivariate Gaussian distributions.
\begin{eqnarray}
    P(\vec{\alpha}) = \mathcal{N}(\vec{0}, \sigma^{2}_{\alpha}I) \label{eq:02}\\ 
    P(\vec{y}|\vec{\alpha}, X) = \mathcal{N}(X \vec{\alpha}, \sigma^2_{y}I) \label{eq:03}
\end{eqnarray}
At this time, the posterior distribution of the surrogate model parameters $\vec{\alpha}$ is computed and given by a multivariate Gaussian distribution, similar to the prior distribution and the likelihood function.
\begin{eqnarray}
    \vec{\alpha}|\vec{y},X  \sim \mathcal{N}(\vec{\mu}, \Sigma) \label{eq:04} \\
    \vec{\mu} = (X^{T}X+ \lambda I)^{-1}X^{T}\vec{y} \nonumber \\    
    \Sigma = \sigma^{2}(X^{T}X+ \lambda I)^{-1} \nonumber \\
    s.t. \,\, \sigma^{2} = \sigma^{2}_{y} \,\, , \,\,  \lambda = \frac{\sigma^{2}_{y}}{\sigma^{2}_{\alpha}} \nonumber
\end{eqnarray}
We sample the surrogate model parameters $\vec{\alpha} \in \mathbb{R}^{p}$ from the multivariate Gaussian distribution described in (\ref{eq:04}).
$\sigma^{2}$ is a hyperparameter indicating the magnitude of fluctuations from the mean vector $\vec{\mu}$ when sampling the surrogate model parameters. $\lambda$ is also a hyperparameter.
Note that $\lambda$ corresponds to the coefficient of the regularization term during ridge regression.

\subsubsection{Construction of acquisition function}
\label{subsubsec: acquisition function}
The acquisition function $f_{acquisition}(\vec{x})$ is constructed in the same QUBO form as the surrogate model, and the next exploration point $\vec{x}^{(new)}$ is proposed by optimizing the acquisition function.
\begin{equation}
    \vec{x}^{(new)} = \underset{\vec{x}}{\argmin} \{f_{acquisition}(\vec{x})\}
\label{eq:05}
\end{equation}
$f_{acquisition}(\vec{x})$ is a function with modified specific parameters from the surrogate model $f_{surrogate}(\vec{x})$ described in \ref{subsubsec: surrogate model}. 
This modification is like a penalty method, designed to ensure that binary vectors with substituent numbers that do not exist at each site do not become the optimal points of the acquisition function. Parameters that are not modified are identical to those in the surrogate model $f_{surrogate} (\vec{x})$.
For example, when six potential substituents can bind at a specific site, representing the 0-indexed substituent numbers in binary requires three bits $(x_{1}, x_{2}, x_{3})$. In this context, $x_{1}x_{2}x_{3} = (000, 001, 010, 011, 100, 101)_{2}$ corresponds to valid substituent numbers from 0 to 5 in decimal. However, each combination $x_{1}x_{2}x_{3} = (110, 111)_{2}$ is equivalent to substituent numbers 6-7 in decimal, rendering them inappropriate as optimal point candidates. To prevent the substituent combinations with substituent numbers 6-7 at this site from being proposed as the optimal points of the acquisition function, we adjust the surrogate model parameters.
In this example, we modify the coefficient of $x_{1}x_{2}$ in the surrogate model function to a positive constant $C$, and the other coefficients are kept the same as in the surrogate model. The next exploration point of the black-box objective function is determined by the optimization of the acquisition function $f_{acquisition}(\vec{x})$.

The search space explored varies greatly depending on how the acquisition function is constructed and how the acquisition function is optimized. As described, our method samples the parameters of the surrogate model and the acquisition function from a probability distribution with variance. The hyperparameter $\sigma^{2}$ indicates the magnitude of the variance. The larger this hyperparameter $\sigma^{2}$ is, the more significant the variance of the acquisition function, potentially allowing for exploration across a broader solution space and avoiding resampling the previously explored points.

\section{Results}
\label{sec: result}
In this section, we describe detailed problem settings and experimental conditions and then show the experimental results obtained by applying our method. In particular, we compare and discuss based on the magnitude of the hyperparameter $\sigma^{2}$. Our discussion centers on two main points of interest in this paper. The first point is whether our method has brought diversity to the proposed substituent combinations. The second point is whether our method has optimized the black-box objective function.
\subsection{Detailed problem settings and experimental conditions}
\label{subsec: detailed settings}
We set the number of substituent binding sites as four, and for convenience in the description, we call each binding site R1, R2, R3, and R4, respectively. The number of possible substituents that can bind at each site is R1: 6, R2: 29, R3: 64, and R4: 64, respectively.
Therefore, the size of the chemical space is calculated as $6 \times 29 \times 64 \times 64 = 712704$.
Moreover, the number of bits necessary to represent the number of each substituent is R1: 3, R2: 5, R3: 6, and R4: 6. 
Consequently, the binary vector $\vec{x}$ dimension is calculated as $N = 3 + 5 + 6 + 6 = 20$. The substituent number at R1 is represented in 0-indexed form using $x_{1}$ to $x_{3}$, similarly, $x_{4}$ to $x_{8}$ represent the substituent number at R2, $x_{9}$ to $x_{14}$ represent the substituent number at R3, and $x_{15}$ to $x_{20}$ represent the substituent number at R4. 
To illustrate with a concrete encoding example, suppose the substituent numbers at each site are R1: 0, R2: 2, R3: 10, and R4: 63. 
In this case, the binary vector $\vec{x}$ would be represented as $\vec{x} = (0,0,0,0,0,0,1,0,0,0,1,0,1,0,1,1,1,1,1,1)$.
We set the hyperparameter $\lambda$ at $10^{-2}$ and the hyperparameter $\sigma^{2}$, which indicates the magnitude of fluctuation from the mean vector $\vec{\mu}$ when sampling surrogate model parameters, to $\{0, 4\times 10^{-3}, 8\times 10^{-3}, 12 \times 10^{-3}\}$.
We set the surrogate model's parameter correction for R1 in the acquisition function as $C = \alpha_{12} = 2 \times max(\vec{\alpha})$ at each after sampling $\vec{\alpha}$.
We used D-Wave Advantage 4.1 as the quantum annealer, setting the annealing time to 2000$\mu$s, and the number of samples is $300$. The quantum adiabatic theorem ensures that it is possible to find the nontrivial ground state at the end of the quantum annealing if the transverse field changes sufficiently slowly \citep{Suzuki2005, Morita2008, Ohzeki2011}. On the other hand, when quantum annealing is carried out on a physical device D-wave quantum annealer, it operates at a finite temperature and is subject to external noise. Due to these factors, the annealing time is often short in many studies. Considering these theoretical and experimental backgrounds, we set the annealing time to be longer in our setting because we observed a tendency for the results to stabilize, possibly due to the effects of ambient temperature.
The number of samples in the initial dataset is 992. 
For comparison as a baseline, we also conducted an experiment where the optimization part of the acquisition function was replaced with random sampling. Due to the nature of this study, which is conducted in the context of actual chemical material exploration, the computational cost of the black-box objective function is exceptionally high, resulting in an experiment of only one instance. We defined one loop as carrying out the following steps (i) through (v), and we performed 20 loops.
\begin{enumerate}[(i)]
 \item \, By sampling the surrogate model parameters $\vec{\alpha}$ from a multivariate Gaussian distribution $\mathcal{N}(\vec{\mu}, \Sigma)$ described in (\ref{eq:04}), construct the surrogate model.
 \item \, Construct the acquisition function by partially correcting the surrogate model parameters as explained in \ref{subsubsec: acquisition function}.
 \item \, Optimize the acquisition function by quantum annealing and select the top 10 points of the acquisition function as the next exploration points for the black-box objective function. In the random sampling used as a baseline, 10 sampling points are randomly selected. Note that at this time, the top 10 points exclude combinations of substituents that are already present in the existing dataset and combinations of substituents that include non-existent substituent numbers, such as substituent numbers 6-7 in R1 and substituent numbers 29-31 in R2, through screening.
  \item \, Take the next exploration points obtained in (iii) as inputs and get outputs, carrying out the evaluation of target material property values, which is the computation of the black-box objective function, through DFT (Density Functional Theory) calculations. 
  The detailed calculation method is described in the Additional Requirements.
 \item \, Append the new samples $\{\vec{x}^{(new)}, y^{(new)}\}$ obtained in (iv) to the existing dataset and return to (i).
\end{enumerate}
    
\subsection{Experimental results}
\label{subsec: detailed results}

\subsubsection{Histogram of substituent numbers in combinations added by end of the experiment}
\label{subsubsec: detailed result1}
\begin{figure}[h]
    \centering
    \includegraphics[width= \textwidth]{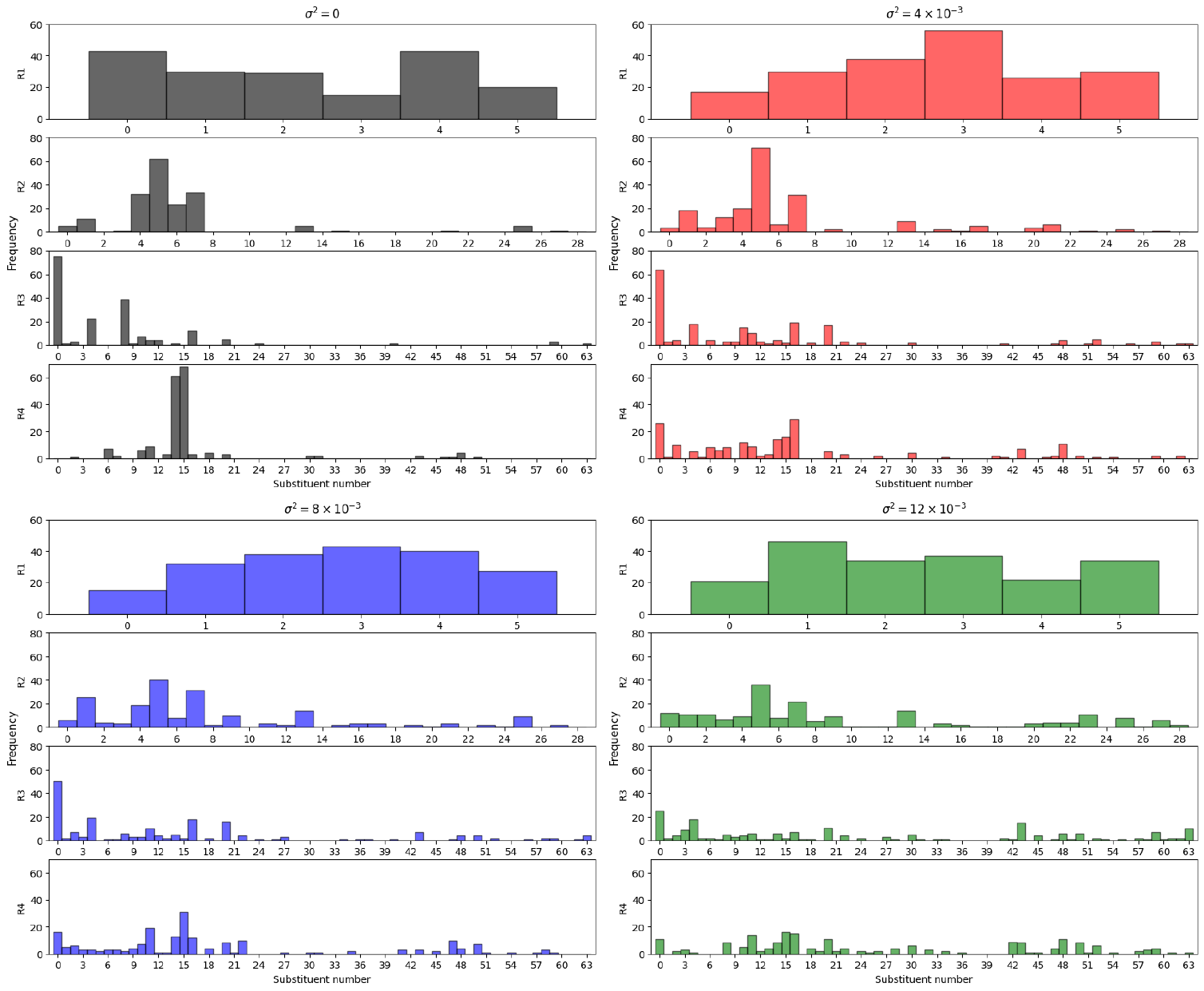} 
        \caption{Histogram of substituent numbers for combinations of substituents added to the dataset by the end of the experiment each $\sigma^{2}$. Top left is $\sigma^2 = 0$, top right is $\sigma^2 = 4 \times 10^{-3}$, bottom left is $\sigma^2 = 8 \times 10^{-3}$, bottom right is $\sigma^2 = 12 \times 10^{-3}$.}
    \label{fig:figure1} 
\end{figure}

We show the histogram of substituent numbers at the binding sites R1, R2, R3, and R4 for the combinations of substituents added to the dataset by the end of the experiment in \textbf{\autoref{fig:figure1}}.
In the case of $\sigma^{2} = 0$, we observed a tendency in R3 and R4 where specific substituent numbers were frequently proposed.
However, as $\sigma^{2}$ increases, it can be observed that diversity is brought into the combinations of substituents proposed for R3 and R4. This difference is particularly pronounced when comparing $\sigma^{2} = 0$ and $\sigma^{2} = 12 \times 10^{-3}$.
From these results, we can infer that we realized the proposal of various combinations of substituents by sampling parameters of the surrogate model and the acquisition function from probability distributions with variance. By sampling parameters from probability distributions with larger variances, the optimal points and the shape of the acquisition function change significantly in each loop. We believe that this approach allowed us to explore the solution space without getting trapped by some specific approximate solutions and without resampling the previously explored points.
\subsubsection{Relationship between the number of loops and the $R^{2}$ of the surrogate model}

\label{subsubsec: detailed result2}
\begin{figure}[h]
    \centering
    \includegraphics[width= 0.8\textwidth]{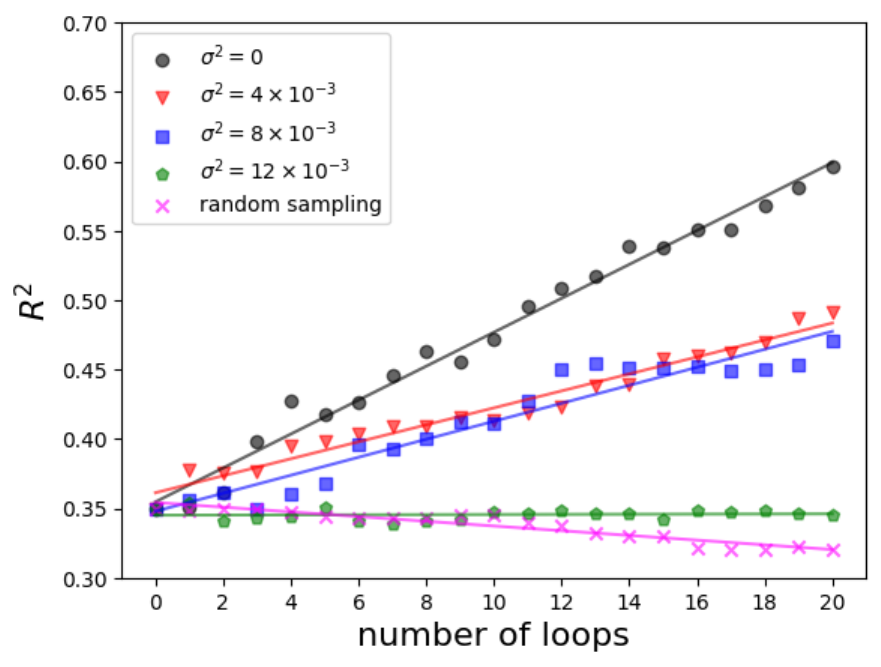} 
    \caption{Relationship between the number of loops and the coefficient of determination $R^{2}$ in the surrogate model each $\sigma^{2}$ and random sampling.}
    \label{fig:figure2} 
\end{figure}

We show the transition of the coefficient of determination $R^{2}$ in the surrogate model at each loop in \textbf{\autoref{fig:figure2}}. The coefficient of determination $R^{2}$ is calculated from the initial dataset sample points, 992 points, and the sample points appended up to each loop. Note that $R^{2}$, plotted in \textbf{\autoref{fig:figure2}}, represents the results of mean-based regression. This result is equivalent to the regression of the maximum a posteriori (MAP) estimation.
As $\sigma^2$ becomes larger, a tendency for $R^{2}$ at each loop to become smaller was observed. We speculate that we can attribute this result to the tendency shown in \textbf{\autoref{fig:figure1}}, where the larger $\sigma^{2}$ is, the more diverse the combinations of substituents that the optimization of the acquisition function proposes become. When $\sigma^2$ is small, $R^2$ improves by fitting to similar input vectors and outputs. However, to improve $R^2$ when $\sigma^2$ is large, it is necessary to fit diverse input vectors and outputs. We speculate that this difficulty is why there was the tendency for the coefficient of determination, $R$, to be smaller when $\sigma^{2}$ is larger.

\subsubsection{Analysis of target material property values}
\label{subsubsec: detailed result3}
\begin{figure}[h]
    \centering
    \includegraphics[width= 0.7 \textwidth]{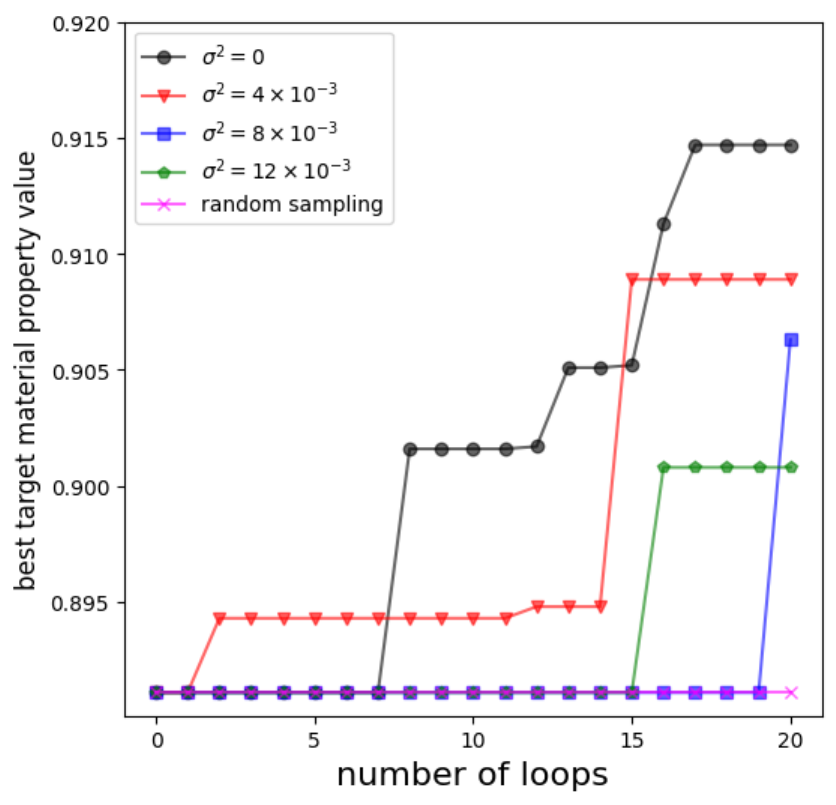} 
        \caption{The transition of the best target material property values in the existing dataset up to each loop.}
    \label{fig:figure3}
\end{figure}

\begin{figure}[h]
    \centering
    \includegraphics[width=\textwidth]{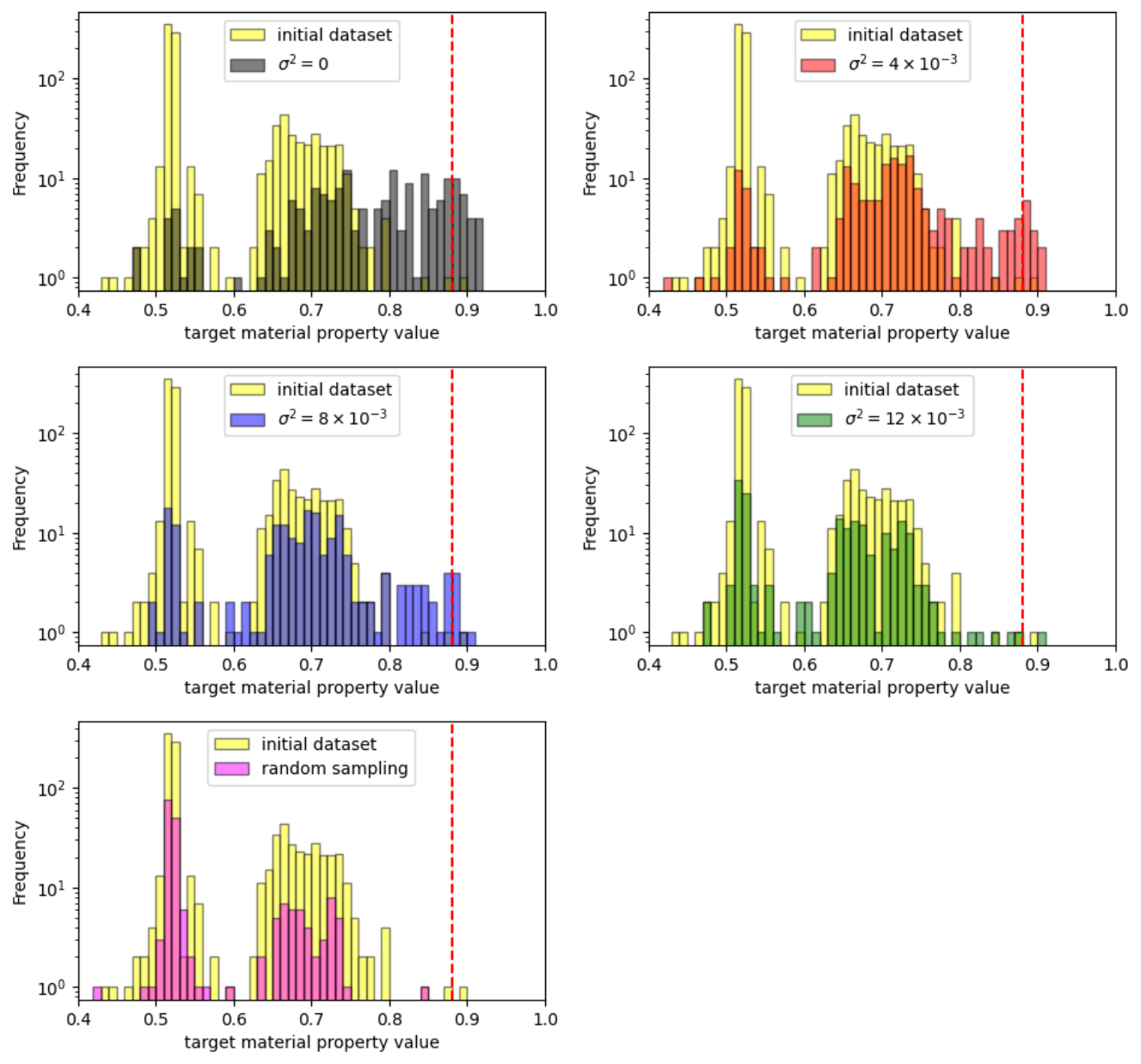} 
        \caption{The histogram of the target material property values for the combinations of substituents in the initial dataset and those added to the dataset by the end of each experiment. The red dotted line shows the cutoff value (0.880), which we defined as a desired target material property value.
        }
    \label{fig:figure4}
\end{figure}

Finally, we show the target material property values evaluated by DFT calculations, corresponding to the combinations of substituents proposed through the optimization of the acquisition function as the next exploration point of the black-box objective function in \textbf{\autoref{fig:figure3}} and \textbf{\autoref{fig:figure4}}.  
In \textbf{\autoref{fig:figure3}}, we plot the transition of the best target material property values in the existing dataset up to each loop. Although we could only experiment once because of the extremely high computational cost of the black-box objective function, in the case of optimizing the acquisition functions, we confirm that it is possible to search for combinations of substituents with higher target material property values than the best value in the initial dataset.
Under the conditions set in this study, using random sampling in the optimization part of the acquisition function, we could not find any combination of substituents that exhibited a property value exceeding the best target material property value in the initial dataset. In \textbf{\autoref{fig:figure4}}, we show the histograms of the target material property values for all combinations of substituents in the initial dataset and those added to the dataset by the end of each experiment.

To reiterate, the objective of black-box optimization in this study was to maximize the target material property value while bringing diversity to the combinations of substituents. Therefore, we listed the combinations of substituents whose target material property values exceeded our criteria of 0.880 or higher in \textbf{\autoref{tab:csvdata1}},\textbf{\autoref{tab:csvdata2}}, \textbf{\autoref{tab:csvdata3}} and \textbf{\autoref{tab:csvdata4}}.
From the perspective of the number of combinations of substituents with property values that exceed our criteria, the number of proposed combinations was the highest at 25 combinations when $\sigma^2 = 0$. 
However, considering the diversity of proposed combinations of substituents, which is one of the aims of this paper, the advantage can be found when $\sigma^{2} \neq 0$. 
Especially in the case of $\sigma^{2} = 4 \times 10^{-3}$, it was possible to discover combinations of substituents with the property values that exceed our criteria, which have the substituent number of R4:0, a combination not discovered in case of $\sigma^{2} = 0$.

\begin{table}[ht]
    \begin{minipage}[t]{0.45\linewidth} 
        \centering
        \caption{Desired target material property value in $\sigma^{2} = 0$}
        \begin{filecontents*}{00000_out.csv}
R1,R2,R3,R4,target material property value
2,5,0,15,0.881
4,7,8,15,0.889
4,5,8,15,0.889
4,5,0,14,0.902
0,5,0,14,0.897
4,6,8,15,0.894
0,5,8,15,0.887
2,5,8,15,0.902
1,5,0,14,0.905
0,7,8,14,0.882
0,6,0,14,0.898
0,7,0,14,0.905
0,5,14,14,0.911
2,5,0,14,0.880
4,6,0,14,0.915
2,7,0,14,0.888
2,7,8,15,0.892
1,6,0,14,0.910
2,6,0,14,0.890
3,6,0,14,0.884
3,5,8,15,0.894
1,7,8,14,0.887
1,6,8,14,0.891
3,7,8,15,0.891
1,7,0,14,0.914
\end{filecontents*}
\csvautotabular{00000_out.csv} 
        \label{tab:csvdata1}
    \end{minipage}
    \hfill
    \begin{minipage}[t]{0.45\linewidth}
        \centering
        \caption{Desired target material property value in $\sigma^{2} = 4 \times 10^{-3}$}
        \begin{filecontents*}{00040_out.csv}
R1,R2,R3,R4,target material property value
0,5,0,15,0.894
3,5,0,14,0.882
3,5,0,15,0.880
2,5,0,15,0.881
4,7,0,0,0.895
3,5,0,0,0.882
1,5,0,14,0.909
0,5,0,14,0.901
3,6,8,15,0.899
3,7,0,0,0.881
2,7,0,0,0.883
\end{filecontents*}
\csvautotabular{00040_out.csv} 
        \label{tab:csvdata2}
        \centering
        \caption{Desired target material property value in $\sigma^{2} = 8 \times 10^{-3}$}
        \begin{filecontents*}{00080_out.csv}
R1,R2,R3,R4,target material property value
2,5,0,15,0.881
2,5,0,14,0.880
4,5,0,15,0.880
4,5,8,15,0.886
0,5,0,14,0.897
1,5,0,14,0.906
\end{filecontents*}
\csvautotabular{00080_out.csv} 
        \label{tab:csvdata3} 
        \centering
        \caption{Desired target material property value in $\sigma^{2} = 12 \times 10^{-3}$}
        \begin{filecontents*}{00120_out.csv}
R1,R2,R3,R4,target material property value
2,7,14,14,0.901
\end{filecontents*}
\csvautotabular{00120_out.csv} 
        \label{tab:csvdata4}
    \end{minipage}
\end{table}

\newpage

\section{Discussion}
\label{sec: discussion}
In this study, we achieved the exploration of diverse approximate solutions in black-box optimization, which has the background of new chemical material discovery, by considering appropriate fluctuations in the parameters of the surrogate model and the acquisition function. 
Although the validity of the result is debatable because of the one-instance experiment, our result indicates that quantum annealing can accelerate the discovery of diverse chemical materials with desired material property values in materials informatics.
More generally, our results demonstrate the advantages and disadvantages of varying the magnitude of the variance when sampling the parameters of the surrogate model from a probability distribution in optimizing a black-box objective function. In this paper, we explored a broader solution space by devising the construction of the surrogate model and the acquisition function. As an alternative approach, we are considering optimizing the acquisition function using a different method from quantum annealing, such as simulated annealing.
Our method in this paper, which encodes combinations of substituents as a binary vector, can be applied even in a more vast chemical space. Future challenges include verifying the performance in such cases and investigating the computational time advantage of using quantum annealing. 

\section{Additional Requirements}
\label{sec: additional requirements}
\subsection{DFT (Density Functional Theory) calculations}
\label{subsec: DFT calculations}
For the proposed substituents by the D-Wave quantum annealer, the energy value of ground and excited states  were calculated by optimizing the geometry based on DFT calculation.DFT calculations were performed using the supercomputer TSUBAME 3.0 with Gaussian16, Revision C.01 software \citep{g16}, with the functional B3LYP and basis functions 6-31G.19 parameters from the DFT calculation were used to reproduce the experimental values.Here, a prediction model was created using random forest regression.

\section*{Conflict of Interest Statement}

Author Yoshihiro Nakao, Takuro Tanaka, and Masami Sako are employed by LG Japan Lab Ltd. Author Masayuki Ohzeki is employed by Sigma-i Co. Ltd. The remaining author declares that the research was conducted in the absence of any commercial or financial relationships that could be construed as a potential conflict of interest.

\section*{Author Contributions}
M.D. developed the experiment method and analyzed the results. Y.N., T.T., and M.S. conducted the numerical experiments related to DFT calculations. M.O. conceived the idea of this study and supervised this research project. All authors discussed the results and contributed to the final manuscript.

\section*{Acknowledgments}
This study used the TSUBAME3.0 supercomputer at Tokyo Institute of Technology.
The authors thank the fruitful discussions with Y. Nishikawa.
Our study receives financial support from the MEXT-Quantum Leap Flagship Program Grant No. JPMXS0120352009, as well as Public\verb|\|Private R\&D Investment Strategic Expansion PrograM (PRISM) and programs for Bridging the gap between R\&D and the IDeal society (society 5.0) and Generating Economic and social value (BRIDGE) from Cabinet Office.


\section*{Data Availability Statement}
The datasets presented in this article are not readily available. Due to the nature of this research, participants did not agree for their data to be shared publicly, so supporting data is not available. The data that support the findings of this study are available from the corresponding author, M.D., upon reasonable request. Requests to access the datasets should be directed to Takuro Tanaka, takuro.tanaka@lgjlab.com.

\bibliographystyle{Frontiers-Harvard} 

\bibliography{preferences}



\end{document}